\documentclass[a4paper,11pt]{article}
\usepackage{pos}
\usepackage{multirow}
\usepackage{graphicx}
\usepackage{setspace}

\title{Spin-Induced Interactions and Heavy-Quark Transport in the QGP}

\author*{Zhanduo Tang}
\author{Ralf Rapp}

\affiliation{Cyclotron Institute and Department of Physics and Astronomy,
Texas A\&M University, College Station, TX
  77843-3366, USA}

\emailAdd{zhanduotang@tamu.edu}
\emailAdd{rapp@comp.tamu.edu}

\abstract{
A previously constructed $T$-matrix approach for studying the quark-gluon plasma (QGP) is improved by incorporating spin-dependent interactions between partons. These interactions arise from the relativistic corrections to the Cornell potential. We first study the vacuum spectroscopy of quarkonia with this potential and find that a significant admixture of a vector component in the confining potential (rather than the previously considered scalar interaction) improves the description of the experimental mass splittings in $S$- and $P$-wave states.
The in-medium potential containing the vector component in the confining interaction is constrained by fitting lattice-QCD results for heavy-quark (HQ) free energies and the equation of state (EoS) computed within in the selfconsistent $T$-matrix  framework. 
We subsequently extract the transport coefficients for charm quarks in the QGP with the improved in-medium potentials. The relativistic corrections to the vector component of the confining potential cause a notable increase in the thermal relaxation rate of charm quarks in the QGP in comparison to previous calculations, especially at high momenta. These results are expected to have significant ramifications for the phenomenology of open heavy-flavor observables at RHIC and the LHC.
}

\FullConference{%
   HardProbes 2023, \\
   26-31 March 2023 \\
   Aschaffenburg, Germany 
}

\begin{document}
\maketitle

\section{Introduction}
Heavy quarks play a key role in the study of both hadron properties in vacuum and the properties of quark-gluon plasma (QGP). The potential between a heavy (charm or bottom) quark ($Q=c,b$) and its antiquark ($\bar Q$) can be tested via quarkonium spectroscopy in vacuum, while heavy-flavor (HF) particles are a powerful probe of the QCD medium as produced in ultra-relativistic heavy-ion collisions (URHICs): The large heavy-quark (HQ) mass, $M_Q$, enables potential approximations and prolongs thermalization times preserving information on the interaction history of HF particles in their finally observed spectra. 
In the present work we advance the thermodynamic $T$-matrix approach to the strongly coupled QGP~\cite{Cabrera:2006wh,Riek:2010fk} by introducing spin-dependent interactions based on a $1/M_Q$ expansion.  
In particular, we introduce a mixed confining potential instead of a purely scalar one~\cite{Mur:1992xv}, which was motivated previously in different contexts~\cite{Szczepaniak:1996tk,Ebert:2002pp}. Its pertinent relativistic corrections will have notable implications for the HQ diffusion coefficient.

\section{T-matrix Approach}
\label{sec_TM}
The thermodynamic $T$-matrix provides a selfconsistent quantum many-body approach to evaluate 1- and 2-body correlation functions. Resummations of the infinite series of ladder diagrams renders it suitable to study both bound and scattering states in a strongly coupled environment. After reducing the 4-dimensional (4D) Bethe-Salpeter into a 3D Lippmann-Schwinger equation~\cite{Brockmann:1996xy}, including a projection on positive-energy states, and a subsequent partial-wave expansion, one
obtains a 1D scattering equation~\cite{Liu:2017qah,ZhanduoTang:2023tdg},
\begin{eqnarray}
\ T_{ij}^{L,a} ( z,p,p')&=&V_{ij}^{{L,a}} (p,p') +\frac{2}{\pi } \int_{0}^{\infty}k^{2}dk V_{ij}^{{L,a}} (p,k) G_{ij}^{0} (z,k)T_{ij}^{L,a} ( z,k,p'), 
\\
G_{i j}^{0}(z, k)&=& \int_{-\infty}^{\infty} d \omega_{1} d \omega_{2} \frac{\left[1 \pm n_{i}\left(\omega_{1}\right) \pm n_{j}\left(\omega_{2}\right)\right]}{z-\omega_{1}-\omega_{2}} \rho_{i}\left(\omega_{1},k\right) \rho_{j}\left(\omega_{2}, k\right), \   
\\
\rho_{i}\left(\omega, k\right)&=&-\frac{1}{\pi} \operatorname{Im}G_{i}(\omega+i \epsilon,k),
G_{i}(\omega,k) =  1/[\omega-\varepsilon_i(k) -\Sigma_i(k)] ,
\end{eqnarray}
with $V_{ij}^{L,a}$: in-medium potential between particle $i$ and $j$ in color ($a$) and angular-momentum $L$ channels; $G_{i j}^{0}$: 2-particle propagator; $G_{i}$ and $\rho_{i}$: 1-particle propagator and spectral function; $n_{i}$: Bose ($+$) or Fermi ($-$) distribution function; $\varepsilon_i=\sqrt{M_{i}^{2}+k^{2}}$; and $p, p'$: moduli of incoming and outgoing momentum in the center-of-mass frame. The single-particle selfenergies, $\Sigma_{i}(k)$, are obtained by closing the $T$-matrix with an in-medium single-parton propagator~\cite{Liu:2017qah,ZhanduoTang:2023tdg}.

\section{Heavy-Quarkonium Spectroscopy in Vacuum}
\label{sec_spectroscopy} 
The vacuum potential used in the $T$-matrix equation is taken as the Fourier transform of a Cornell potential, including relativistic corrections~\cite{Riek:2010fk,Lucha:1995zv}.
Specifically, we account for spin-orbit and spin-spin interactions resulting in
\begin{equation}
\begin{aligned}
&V_{Q\bar{Q}}\left(\mathbf{p}, \mathbf{p}^{\prime}\right)=\mathcal{R}_{Q\bar{Q}}^{vec} V^{vec}\left(\mathbf{p}-\mathbf{p}^{\prime}\right)+\mathcal{R}_{Q\bar{Q}}^{sca}V^{sca}\left(\mathbf{p}-\mathbf{p}^{\prime}\right)+\mathcal{R}_{Q\bar{Q}}^{spin}[V^{LS}(\mathbf{p}-\mathbf{p'})+V^{SS}(\mathbf{p}-\mathbf{p'})+V^{T}(\mathbf{p}-\mathbf{p'})],
\end{aligned}
\label{eq_V}
\end{equation}
where $V^{vec}$ ($V^{sca}$) is the static vector (scalar) potential. The spin-dependent potentials are given by
\begin{eqnarray}
{V^{LS}}(r) &=&\frac{1}{2M_Q^{2}r}\left \langle \mathbf{L\cdot S} \right \rangle \left (3\frac{d}{dr}V^{vec}(r)-\frac{d}{dr}V^{sca}(r)\right),\nonumber
{V^{SS}}(r) =\frac{2}{3M_Q^{2}}\left \langle \mathbf{S_1\cdot S_2} \right \rangle \Delta V^{vec}(r),\nonumber\\
{V}^{T}(r) &=&\frac{1}{12M_Q^{2}}\left \langle S_{12} \right \rangle \left (\frac{1}{r}\frac{d}{dr}V^{vec}(r)-\frac{d^2}{dr^2}V^{vec}(r)\right) \ .
\label{eq_VLSSST}
\end{eqnarray}
A Gaussian smearing is applied to the $\delta$-function in the $SS$ component, $\tilde{\delta}(r)=\left(\frac{b}{\sqrt{\pi}}\right)^3 e^{-b^2 r^2}$~\cite{Hong:2022sht}. We set $b=10$, having verified that the $SS$ interaction reaches saturation in quarkonium spectroscopy for $b>10$. The expectation values, $\left \langle \mathbf{L\cdot S} \right \rangle$, $\left \langle \mathbf{S_1\cdot S_2} \right \rangle$, and $\left \langle S_{12} \right \rangle$, follow the standard form described in Ref.~\cite{Lucha:1995zv}, and the relativistic correction factors in Eq.~(\ref{eq_V}) are given 
by~\cite{Cabrera:2006wh,Riek:2010fk,Liu:2017qah}
\begin{equation}
\begin{aligned}
&\mathcal{R}_{Q\bar{Q}}^{vec}=\sqrt{1+\frac{p^{2}}{\varepsilon _{Q}(p)\varepsilon_{\bar{Q}}(p)}}\sqrt{1+\frac{p'^{2}}{\varepsilon _{Q}(p')\varepsilon _{\bar{Q}}(p')}},
\mathcal{R}_{Q\bar{Q}}^{sca}=\mathcal{R}_{Q\bar{Q}}^{spin}=\sqrt{\frac{M_{Q}M_{\bar{Q}}}{\varepsilon_Q(p)\varepsilon_{\bar{Q}}(p)}}\sqrt{\frac{M_{Q}M_{\bar{Q}}}{\varepsilon_Q(p')\varepsilon_{\bar{Q}}(p')}}.
\label{eq_R correction new}
\end{aligned}
\end{equation}
The Lorentz structure for color-Coulomb potential, $V_\mathcal{C}=-\frac{4}{3}\frac{\alpha_s }{r}$, is entirely vector, and a common assumption for the confining one, $V_\mathcal{S}=\sigma r$, is to be entirely scalar, {\em i.e.},  $V^{vec}=V_\mathcal{C}$ and $V^{sca}=V_\mathcal{S}$. The coupling constant, $\alpha_s=0.27$, and string tension, $\sigma=0.225$ $\textup{GeV}^2$, are fitted to the vacuum free energy from lQCD data~\cite{Bazavov:2018wmo}, with a string breaking distance of $r_{SB}$=1\,fm. As stated in the introduction, there are indications that the confining potential exhibits a combination of vector and scalar Lorentz structures, {\em i.e.}, 
$V^{vec}=V_\mathcal{C}+(1-\chi)V_\mathcal{S}$ and $V^{sca}=\chi V_\mathcal{S}$, rather than being purely scalar. The crucial parameter in this context is the mixing coefficient, denoted by $\chi$, defined such that $\chi=1$ corresponds to a purely scalar potential, while values below one admix a vector component. 

\begin{figure}[htbp]
\begin{minipage}[b]{1.0\linewidth}
\centering
\includegraphics[width=0.9\textwidth]{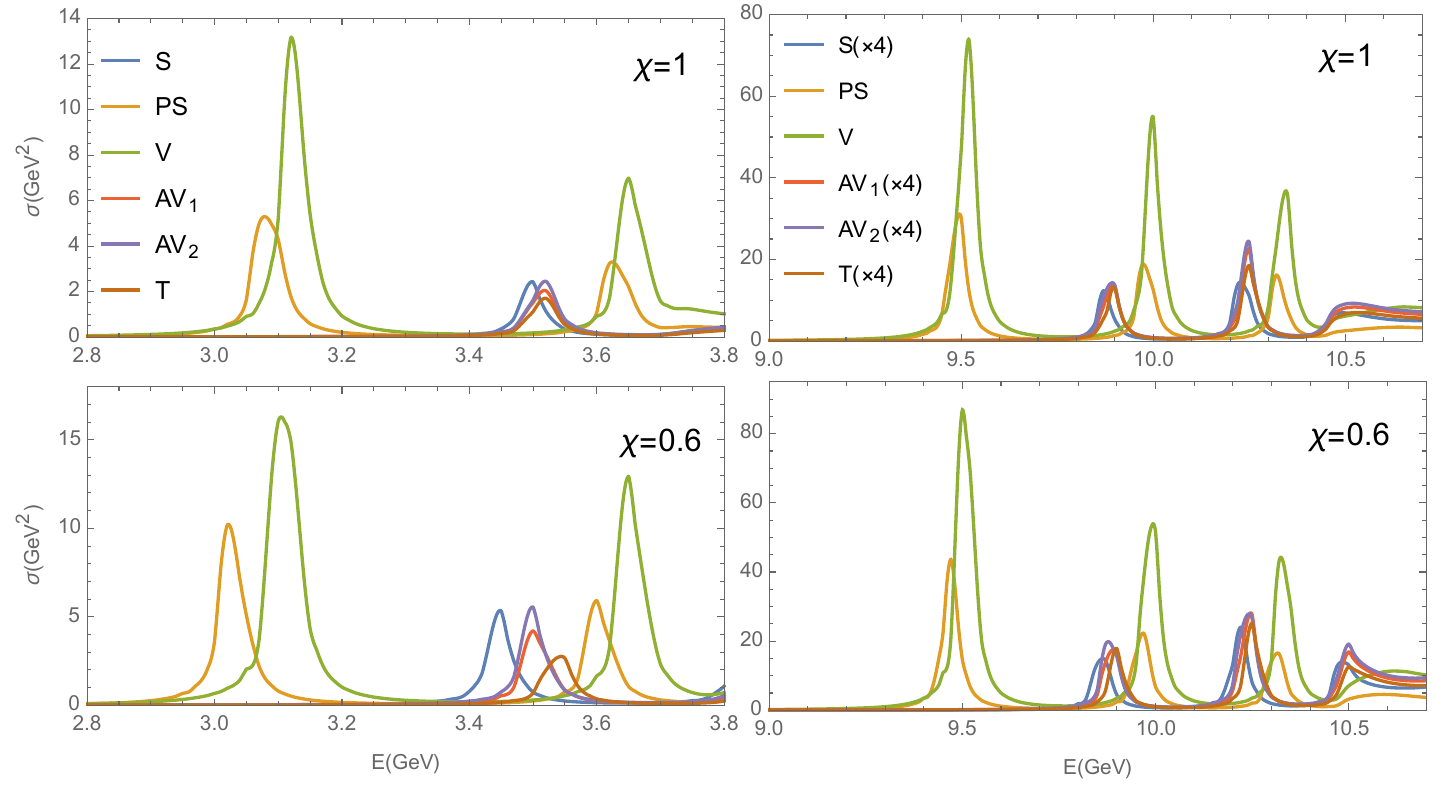}
\end{minipage}
\caption{The vacuum charmonium (left panels) and bottomonium (right panels) spectral functions in different mesonic channels with mixing coefficient $\chi=1$ (upper panels) and $\chi=0.6$ (lower panels).} 
\label{fig_sf-vac}
\end{figure}

The quarkonium masses are extracted from the peak values in the pertinent meson spectral functions, $\sigma(E)= - \frac{1}{\pi}\operatorname{Im}G(E+i\epsilon)$, obtained from the correlation functions, $G(E)=G_0(E)+\Delta G(E)$. In the CM frame they are given by
\begin{equation}
\begin{aligned}
& G_{0}(E)&=&N_{f}N_{c}\int \frac{d^3p}{(2\pi )^3}\mathcal{R}_{Q\bar{Q}}^{sca} a_{non}(p) G_{Q\bar{Q}}^0(E,p), \\
&\Delta G(E)&=&\frac{N_fN_c}{\pi ^3}\int \mathcal{R}_{Q\bar{Q}}^{sca} dp p^2  G_{Q\bar{Q}}^0(E,p)\int dp' p'^2 G_{Q\bar{Q}}^0(E,p')[a_0(p,p')T_{Q\bar{Q}}^{0}+a_1(p,p')T_{Q\bar{Q}}^{1}] 
\label{eq_G0}
\end{aligned}
\end{equation}
($N_f=1$, $N_c=3$). The leading-order in $1/M_Q$ values of the $a_{0,1,non}$ for the scalar (S), pseudoscalar (PS), vector (V), axial-vector (AV) and tensor (T) mesonic channels are given in Ref.~\cite{ZhanduoTang:2023tdg}. The resulting quarkonium spectral functions clearly show the improvement in the spin splittings once the vector component in the confining force is introduced, especially for charmonia, see Fig.~\ref{fig_sf-vac}.
We also found that, for $\chi=0.6$, the experimental splitting between $D$ and $D^*$ mesons is well reproduced. This is particularly relevant for calculating the HQ transport coefficients which are based on HQ interactions with the light partons in the QGP as discussed in the following section.
\section{Charm-Quark Transport Coefficients in the QGP}
\label{sec_transport}
\begin{figure}[htbp]
\begin{minipage}[b]{1.0\linewidth}
\centering
\includegraphics[width=1\textwidth]{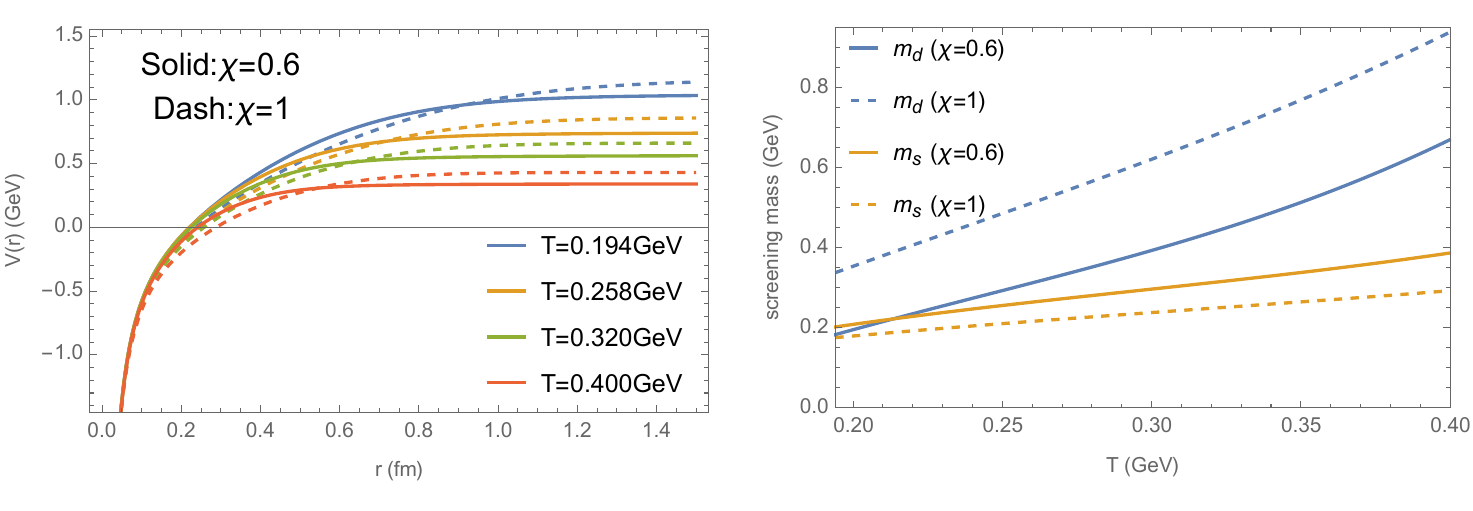}
\end{minipage}
\caption{Left panel: The in-medium static potentials for $\chi=0.6$ (solid) and 1 (dashed) at different temperatures. Right panel: The temperature dependence of the 
screening masses, $m_d$ (blue) and $m_s$ (orange), for $\chi=0.6$ (solid) and 1 (dashed).} 
\label{fig_free energy}
\end{figure}

The in-medium potentials, $V(r,T) = -\frac{4}{3} \alpha_{s} [\frac{e^{-m_{d} r}}{r} + m_{d}] -\frac{\sigma}{m_s} [e^{-m_{s} r-\left(c_{b} m_{s} r\right)^{2}}-1]$, used for computing the charm-quark transport coefficients are constrained by lQCD data for HQ free energies~\cite{Bazavov:2012fk} and the
equation of state~\cite{HotQCD:2014kol} and are shown in Fig.~\ref{fig_free energy}. 

At low momenta, the $T$-matrices for $\chi=0.6$ are smaller than those for $\chi=1$ due to a stronger screening effect in the confining potential; however, for $p_{cm}\gtrsim$0.5 GeV, the amplitudes for $\chi=0.6$ exceed those for $\chi=1$, which is attributed to the harder dependence on 3-momentum of the confining potential from relativistic effects.
Consequently, the pertinent parton spectral functions, $\rho_{q,g,c}$, still exhibit large widths at low momenta, while for higher momenta their decrease is much less pronounced for $\chi=0.6$ compared to $\chi=1$.  



The thermal charm-quark relaxation rate in the QGP can be calculated as~\cite{Liu:2018syc}
\begin{equation}
\begin{aligned}
A(p)=& \sum_{i} \frac{1}{2 \varepsilon_{c}(p)} \int \frac{d \omega^{\prime} d^{3} \mathbf{p}^{\prime}}{(2 \pi)^{3} 2 \varepsilon_{c}\left(p^{\prime}\right)} \frac{d \nu d^{3} \mathbf{q}}{(2 \pi)^{3} 2 \varepsilon_{i}(q)} \frac{d \nu^{\prime} d^{3} \mathbf{q}^{\prime}}{(2 \pi)^{3} 2 \varepsilon_{i}\left(q^{\prime}\right)} \delta^{(4)} \frac{(2 \pi)^{4}}{d_{c}} \\
& \times \sum_{a, l, s}|M|^{2} \rho_{c}\left(\omega^{\prime}, p^{\prime}\right) \rho_{i}(\nu, q) \rho_{i}\left(\nu^{\prime}, q^{\prime}\right) \left[1-n_{c}\left(\omega^{\prime}\right)\right] n_{i}(\nu)\left[1 \pm n_{i}\left(\nu^{\prime}\right)\right] (1-\frac{\mathbf{p}\cdot\mathbf{p'}}{\mathbf{p}^2}) \ ,
\end{aligned}
\label{eq_A(p)}
\end{equation}
The summation, $\sum_{i}$, is over all light quarks and gluons. All partons are treated with off-shell spectral functions except for the incoming charm quark. The heavy-light scattering amplitudes, $|M|^2$, are directly related to the $T$-matrices, cf.~Ref.~\cite{Liu:2018syc}
 
\begin{figure}[htbp]
\begin{minipage}[b]{1.0\linewidth}
\centering
\includegraphics[width=1\textwidth]{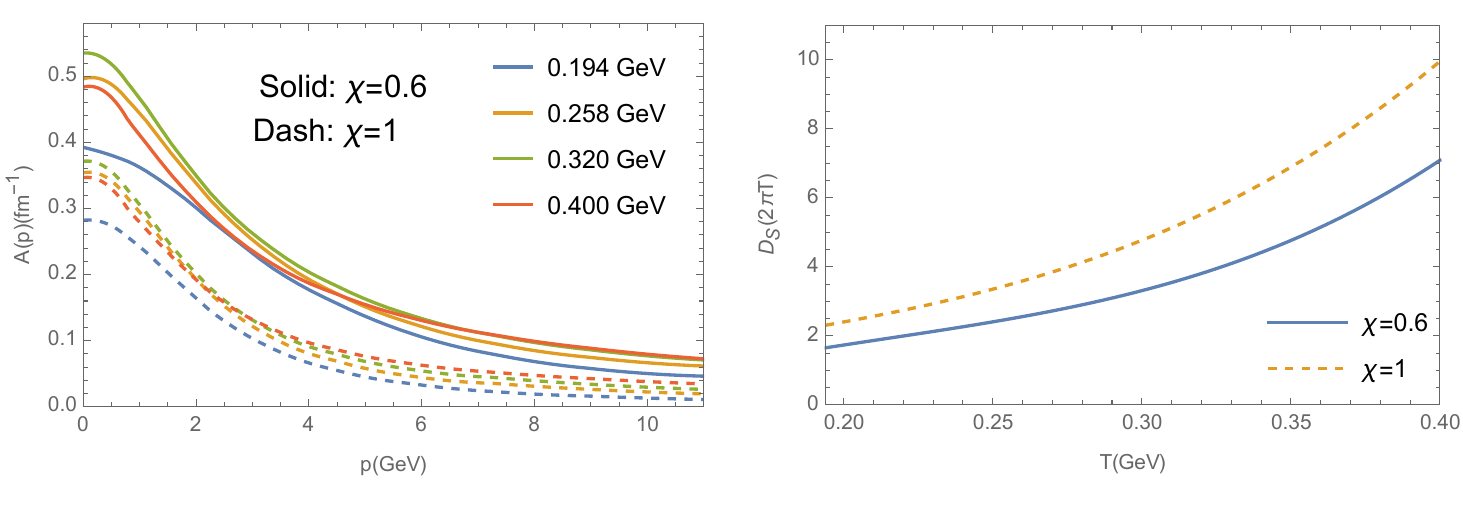}
\end{minipage}
\caption{The charm-quark relaxation rate at different temperatures (left panel) and spatial diffusion coefficient (right panel) for $\chi=0.6$ (solid) and 1 (dashed).} 
\label{fig_Ap}
\end{figure}

The resulting relaxation rate exhibits notable enhancement at low momenta due the vector component in the confining potential, see Fig.~\ref{fig_Ap} left. However, the more significant impact is observed at higher momenta due to the its harder momentum dependence from relativistic effects.
The spatial diffusion coefficient, $D_s=T/(M_c A(p=0))$, scaled by the inverse thermal wavelength, $2\pi T$, is slightly smaller for $\chi=0.6$ compared to $\chi=1$ due to the non-zero average momenta in the CM of the heavy-light interactions, cf.~right panel in Fig.~\ref{fig_Ap}.
\section{Conclusions}
\label{sec_concl}
We have extended the thermodynamic $T$-matrix approach to incorporate  $1/M_Q$ corrections via spin-dependent interactions between heavy quarks. With an additional admixture of a Lorentz-vector component in the confining potential, a significantly improved description for vacuum charmonium spectroscopy has been achieved. Its relativistic corrections also lead to harder scattering amplitudes and pertinent parton spectral functions in the QGP.
Employing these to compute the thermal relaxation rate of charm quarks results
in an enhancement that becomes more pronounced toward higher 3-momenta, compared to a purely scalar confining potential (along with a corresponding smaller diffusion coefficient). These findings are encouraging as they are likely to improve the phenomenology heavy-flavor diffusion in heavy-ion collisions at RHIC and the LHC based on nonperturbative interactions.
\acknowledgments
This work has been supported by the U.S. National Science Foundation under grant nos. PHY-1913286 and PHY-2209335, and by the U.S. Department of Energy, Office of Science, Office of Nuclear Physics through the Topical Collaboration in Nuclear Theory on \textit{Heavy-Flavor Theory (HEFTY) for QCD Matter} under award no.~DE-SC0023547.

\bibliographystyle{unsrt} 
\bibliography{ref}


\end{document}